\documentclass[pre,floatfix,twocolumn,showpacs,a4paper]{revtex4}
\usepackage{amsmath}
\usepackage{amsfonts}
\usepackage{amssymb}
\usepackage{calrsfs}
\usepackage{mathrsfs}
\usepackage{epsfig}
\usepackage{graphicx}
\usepackage{dcolumn}
\usepackage{color}
\usepackage{url}

\usepackage{hyperref}

\begin {document}

\title
{
Network similarity and statistical analysis of earthquake seismic data
}
\author
{
Krishanu Deyasi$^{1}$, Abhijit Chakraborty$^{2}$ and Anirban Banerjee$^{1,3}$ 
}
\affiliation
{
\begin {tabular}{c}
$^1$ Department of Mathematics and Statistics, Indian Institute of Science Education and Research Kolkata, Mohanpur-741246, India \\
$^2$ Graduate School of Simulation Studies, University of Hyogo, Kobe 650-0047, Japan \\
$^3$ Department of Biological Sciences, Indian Institute of Science Education and Research Kolkata, Mohanpur-741246, India \\
\end{tabular}
}

\begin{abstract}
We study the structural similarity of earthquake networks constructed from seismic catalogs of different geographical regions.
A hierarchical clustering of underlying undirected earthquake networks is shown using Jensen-Shannon divergence in graph spectra. The directed 
nature of links indicates that each earthquake network is strongly connected, which motivates us to study the directed version statistically. Our statistical analysis
of each earthquake region identifies the hub regions. We calculate the conditional probability of the forthcoming occurrences of earthquakes in each region.
The conditional probability of each event has been compared with their stationary distribution.

\end{abstract}

\maketitle
\section {Introduction}
 Earthquakes are one of the most devastating natural calamities that can shatter human civilization in large extent. 
Naturally, scientists are investigating this phenomena in great detail. The most robust empirically established facts 
in the phenomenology of earthquakes are the Gutenberg-Richter law~\cite{GR} and the Omori law~\cite{Omori}. While the Gutenberg-Richter law 
expresses the relationship between frequency and magnitude of tremors in a region, the Omori law describes the temporal rate of decay of aftershocks.
Different models have been proposed to study the phenomena of earthquakes. One well-known example is the Burridge-Knopoff
model that demonstrates the statistical properties of earthquakes using friction on a fault
surface as a stickslip process~\cite{BK}.  It is considered that large faults in the Earth's crust are formed due to the
action of plate tectonic forces.

Plate tectonic theory, which is based on Alfred Wegener's continental drift theory~\cite{Cox09}, is considered to
be the fundamental theoretical framework in the field of Earth Science and plays the pivotal role to explain tremors. 
This theory states that Earth's outer shell, the lithosphere is divided into several rigid pieces, called plates.
There are mainly eight major plates: African, Antarctic, Eurasian, North American, South American, Pacific, and Indo-Australian. 
When a pair of plates move with respect to each other they do not deform internally rather they deform along their edges and create 
earthquakes and volcanoes along the edges of the plates.

Recently a complex network approach has been applied to observe the universal features of the earthquake phenomena from seismic catalogs. 
Based on the work of Bak {\it et.al.}~\cite{Bak}, Baiesi and Paczuski, using a correlation metric between a pair of
events, have constructed a network in which tremors are the nodes and a pair of nodes are linked if the correlation 
between them is higher than a certain threshold~\cite{Baiesi2004, Davidsen2005}. It is shown that the earthquake network
exhibits a scale-free nature with highly heterogeneous degree distribution characterized by a power law. 
In another study, Abe and Sazuki have constructed the network
using a grid, covering the entire earthquake events over a region~\cite{Abe1, Abe2, Abe3, Abe4, Abe5, Abe6, Abe7, Abe8}.
A cell of the grid is considered to be a
node if at least one epicenter occurs within the cell, and the cell size is a tunable parameter for the model. A pair of nodes
is connected by a link if two successive events occur on those two nodes. Using this network model, they found various
robust features of the earthquake network. They have shown that earthquake networks exhibit a power-law degree
distribution~\cite{Abe1} small-world phenomena~\cite{Abe2}, assortativity~\cite{Abe3} and scaling in local clustering~\cite{Abe8}.
Subsequently, different structural properties of weighted earthquake network have also been studied extensively in~\cite{Abhijit15}.

Although universal features of earthquake networks of different regions have been studied extensively~\cite{Baiesi2004,Abe1, 
Abe2, Abe3, Abe4, Abe5, Abe6, Abe7, Abe8}, but no study has thus for been devoted to capturing the similarity and dissimilarity 
between earthquake networks of different regions. A study of hierarchical clustering~\cite{Johnson1967} of 
earthquake networks will be very useful in this regard. Hierarchical clustering is usually represented by a dendrogram that   
pictorially shows the similar entities are clustered together in groups. Quantitatively, the similarity between networks could be measured based
on different properties, viz., Euclidean distance, structural properties and dynamical behavior.  
 
Primarily, earthquake networks are directed sequence of consecutive events, so analyzing the directed structure of the earthquake network can give more
information than the underlying structure of the earthquake network~\cite{Abe05}. Not only the directed structure of earthquake network, studying the directed sequence of earthquake events
can also give us important insight, such as, prediction of consecutive earthquakes.   
 
In this paper, we divide our study of earthquake data into two parts. In the first part, we follow the method of Abe
and Suzuki to construct the earthquake network as it is connected to a universal law~\cite{Abe9}. 
The similarity between each pair of spectral probability functions  of eleven earthquake networks is measured using a probabilistic
measure viz., Jensen-Shanon divergence. Using the similarity distances, the hierarchical clustering between earthquake networks is
shown as a dendrogram. The hierarchical clustering of earthquake networks reveals the similar or dissimilar nature
of different earthquake regions based on their position on different tectonic plates. We also measure the frequency of earthquake 
events on a node and identify the earthquake prone locations for different regions. In the second part, we 
find the pair of regions where consecutive earthquake events have occurred with relatively higher frequency. For this, we calculate 
the conditional probability between two successive events. We also compare the conditional probability of two consecutive earthquakes with their
stationary conditional probability to predict the occurrence of an earthquake event at a node consecutively after one occurs at a certain node. 

\begin{table*}[top] 
\begin {center}
\begin{tabular}{|l|l|l|l|l|l|l|l|} \hline
  Region       & Period             &  $\theta_{min}$    & $\theta_{max}$      & $\phi_{min}$    &  $\phi_{max}$   & $n$          & $(L_{lat}L_{lon})^{1/2}$ \\ \hline
  SC           & 1973 - 2011        &   32               &   37                & -122            & -114            & 572601       & 638.33  \\ 
  NC           & 1972 - 2014        & 35.301             & 42.223              & -126.145        & -116.643        & 99998        & 796.21  \\
  JAP          & 1985 - 1998        &  25.73             & 47.964              & 126.43          &  148.0          & 200910       & 2178.01  \\
  CAN          & 1985 - 2015        &  41.01             &  83.54              & -149.89         & -41.83          & 83404        & 5140.65  \\ 
  IRAN         & 1990 - 2014        &  20.650            &  44.490             &  40.000         &  69.590         &  21057       &  2710.74  \\                    
  GR           & 1970 - 2012        & 33.100             & 43.680              & 14.810          & 35.030          & 139126       & 1439.64  \\                
  NM           & 1974 - 2014        & 26.716             & 43.780              & -98.880         & -74.540         & 10425        & 2047.57  \\         
  BI           & 1970 - 2014        & 49.009             & 63.000              & -10.904         & 5.000           & 9754         & 1240.08  \\ 
  AUS          & 1985 - 2015        & -43.78             & -10.16              & 110.50          & 154.89          & 24126        & 4054.84  \\
  SZ           & 1951 - 2008         & 45.40              & 48.30               &  5.60           &  11.10          & 15767        & 367.20   \\
  NZ           & 2000 - 2015        & -49.18             & -32.28              &  -179.99        &  180.00         & 319353       & 7547.32  \\ \hline
     \end{tabular}
     \label{table1}
\caption{ Specification of the parameters of different earthquake catalogs. 
All angles are measured in degree and $(L_{lat}L_{lon})^{1/2}$ is measured in kilometer.\\
{\bf SC}:  Southern California Earthquake Data Center, \url {http://www.data.scec.org/}\\
{\bf NC} : USGS NCSN catalog,  \url {http://quake.geo.berkeley.edu/ncedc/catalog-search.html} \\
{\bf JAP}: Japan University Network Earthquake Catalog, \url {http://wwweic.eri.u-tokyo.ac.jp/db/junec/} \\
{\bf CAN}: Canada's National Earthquake Database, \url {http://earthquake.usgs.gov/earthquakes/eqarchives/} \\
{\bf IRAN}: International Institute of Earthquake Engineering and Seismology,  \url {http://www.iiees.ac.ir} \\
{\bf GR} : Institute of Geodynamics, \url {http://www.gein.noa.gr/services/cat.html} \\
{\bf NM} : Centre for Earthquake Research and Information, Unv. of Memphis, \url {http://www.ceri.memphis.edu/seismic/catalogs/} \\
{\bf BI} : BGS Earthquake Database, \url {http://www.earthquakes.bgs.ac.uk/earthquakes/dataSearch.html} \\
{\bf AUS}: Geoscience Australia - Earthquake Database, \url {http://www.ga.gov.au/earthquakes/searchQuake.do} \\
{\bf SZ}:  Swiss Seismological Service, \url {http://hitseddb.ethz.ch:8080/ecos09/query\_sum} \\
{\bf NZ} : GeoNet – the official source of geological hazard information for New Zealand, \url {http://quakesearch.geonet.org.nz/} \\
}
\end {center}
\end {table*}

 
\section {Data}   

\begin{figure}[top]
\begin{center}
\includegraphics[width=7.0cm]{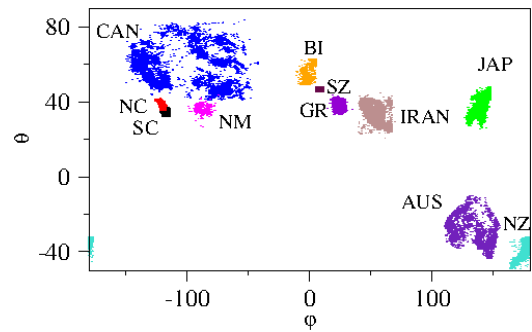}
\end{center}
\caption{(Color online)
Data points representing the positions of the epicenters of the earthquakes events.
 Different regions are indicated with different colors.}
\label{figure1}
 \end{figure}

    We have analyzed eleven distinct earthquake catalogs for different parts of the world, namely the Southern California Earthquake Data Center 
   catalog (SC), Northern California Earthquake Catalog (NC), Japan University Network Earthquake catalog (JAP), Canada's National Earthquake Database catalog 
   (CAN), International Institute of Earthquake Engineering and Seismology - Iran catalog (IRAN), Institute of Geodynamics -Greece catalog (GR),
   Center for Earthquake Research and Information - New Madrid catalog (NM) and British Geological Survey Earthquake Database around the British Isles catalog (BI),
   Geoscience Australia catalog(AUS),  Swiss Seismological Service catalog (SZ) and GeoNet - the official source of geological hazard information for New Zealand catalog (NZ).
   Each catalog contains the geographical positions of the epicenters, specified by their latitudes 
   ($\theta$) and longitudes ($\phi$) and the exact occurrence times of the tremors. The positions of the
   epicenters of the different earthquake regions  are shown in Fig. \ref{figure1}. All the related parameters are mentioned in Table \ref{table1}. 
   The minimum and maximum values of the latitude-longitude coordinates, i.e., ($\theta_{min}, \theta_{max}$) and ($\phi_{min}, \phi_{max}$)
   characterize the extent of a earthquake region. The entire earthquake region is discretized into a two dimensional grid,
   following the approach in~\cite {Abe1}. In this method, the cell size $L$ is the parameter of the model. We use the definition~\cite {Abe7}  of 
   the dimensionless cell size parameter $\ell = L/(L_{lat}L_{lon})^{1/2}$.  Here,  the total extent along the north-south and the east-west directions 
   of the entire earthquake region are $L_{lat}$ and $L_{lon}$ respectively. The North-South distance between
   $(\theta_i, \phi_i)$ and $(\theta_{min}, \phi_{min})$ is $d_{NS} = R(\theta_i -\theta_{min})$
   and the East-West distance is $d_{EW} = R(\phi_i -\phi_{min})cos\theta_{av}$, where the 
   radius of the earth is $R=6370$ Km and $\theta_{av}$ is $(\theta_{min} + \theta_{max})/2$.       
   The parameter $n$ in Table~\ref{table1} represents the total number of earthquake events in the catalog.

\section {Construction of the earthquake network}
We use the method of Abe and Suzuki~\cite {Abe1} to construct the earthquake network.
In this method, the entire earthquake region has been discretized into a two dimensional
rectangular grid with the dimensionless cell size $\ell$ as tunable parameter.
Here, a cell is considered to be a node if at least one earthquake has its epicenter within this cell. 
A pair of nodes is connected by a link if and only if at least one pair of successive events occur 
whose epicenters are located within these two cells. The weight $w_{ij}$ of a link connecting
two distinct nodes $i$ and $j$, is the total number of consecutive events between them~\cite{Abhijit15}.
The strength of the $i$-th node is defined as $s_i=\sum_j w_{ij}$.
In general, there can be self loops of the nodes in the network, but we do not consider those for network construction.

\section {Results}  

\begin{figure}[top]
\begin{center}
\includegraphics[width=7.0cm]{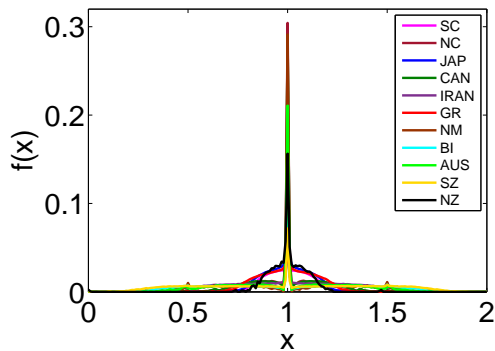}
\end{center}
\caption{(Color online)
 Spectral plot of the earthquake networks using $\ell = 0.01$.  
 The spectral probability function $f(x)$ has been plotted as a collection of the eigenvalues $\lambda_i$ by convolving with a Gaussian kernel with $\sigma = 0.005$.}
 \label{figure2}
\end{figure}
 We study the spectral plots of the normalized graph Laplacian operator $(\Delta)$ of the earthquake networks.
 The normalized graph Laplacian operator $(\Delta)$~\cite{BanerjeeJost2008a} of an unweighted and undirected graph is defined as:
\begin{equation}
(\Delta)_{ij} =
\begin{cases}
1 & \text{if } i=j,
\\
-\frac{1}{k_i} & \text{if $i$ and $j$ are neighbors},
\\
0 & \text{otherwise } 
\end{cases}
\end{equation}
 where $k_i$ is the degree of the $i$-th node. Note that this operator is similar to the operator studied in~\cite{Chung}.

 The spectrum of a network has been convolved with a kernel $g(x,\lambda)$ to compute the graph spectra.
 After convolution, we get the function
\begin{equation}      
 f(x) = \int g(x,\lambda) \sum_k \delta(\lambda,\lambda_k)d\lambda = \sum_k g(x,\lambda_k)      
\end{equation} 
 Evidently,
\begin{equation}
0 < \int f(x)dx < \infty.
\end{equation}
Here, we use a Gaussian kernel $1/\sqrt{2\pi\sigma^2} exp(-(x-m_x)^2/2\sigma^2)$ with $\sigma=0.005$.
Now $f(x)$ can be written as $f(x)=\sum_{\lambda_i} 1/0.005 \sqrt{2\pi} exp(-(x-\lambda_i)^2/0.00005)$. 
 
 The spectral probability functions $f(x)$ of the earthquake networks for different earthquake regions  are 
 shown in Fig. \ref{figure2} for $\ell = 0.01$. It is observed that each function has a peak at $x=1$ and the degree of the peakedness
 of different distributions vary significantly. The nature of $f(x)$ is very different from the oval shaped spectral probability functions
 of Erd\H os-R\'enyi random network~\cite{BanerjeeJost2009, Lange2014}. 
 
\subsection{Part A: Structural similarity in earthquake networks} 
 We now measure the pairwise similarity of the spectral probability functions using the Jensen-Shannon (JS) divergence~\cite{Endres2003}
 for a pair of probability distribution $p_1(x)$ and $p_2(x)$ of a discrete random variable $x$ namely 
$$JS(p_1,p_2)=\frac{1}{2}KL(p_1,p)+\frac{1}{2}KL(p_2,p),$$
where $p=(p_1+p_2)/2$. 
The above JS divergence is defined in terms of KL divergence: 
$$ KL(p_1,p_2)=\sum_{x \in X}p_1(x)log\frac{p_1(x)}{p_2(x)},$$
The JS divergence has many advantages over the KL divergence.
For instance, it is symmetric and it is also defined when any one of the probability
distribution ($p_1$ or $p_2$) is zero. It is known that the square root of JS divergence is a metric~\cite{Lin1991}.

The structural distance $D(\Gamma_1,\Gamma_2)$ between a pair of networks $\Gamma_1$ and $\Gamma_2 $ is defined as~\cite{Anirban, Deyasi15} :
\begin{equation}
D(\Gamma_1,\Gamma_2)=\sqrt{JS(f_1,f_2)}
\end{equation}

  The structural distances $D(\Gamma_1,\Gamma_2)$ between different earthquake networks are shown in Table \ref{table2}. 
 The distance matrix $D(\Gamma_i,\Gamma_j)$ is used for hierarchical clustering of earthquake networks of different regions.
 The hierarchical clustering method~\cite{Johnson1967} consists of four steps:
(i) Each individual network is treated as a cluster. 
(ii) A pair of clusters, separated by the shortest distance are merged and form a single cluster together.
(iii) The distances between all pair of clusters are computed.
(iv) Steps (ii) and (iii) are repeated until all the individual networks form a single cluster. 

In step (iii) the distance between clusters can be calculated in different ways.
For instance, complete linkage clustering conside the maximum distance is considered between the pair of clusters.
Similarly, it is called the single linkage  clustering (average linkage clustering) if the minimum distance
(average distance) is considered between the pair of clusters. Note that the hierarchical clustering formed
using the complete linkage algorithm is equivalent to a network in which a pair of nodes are linked if the distance 
between them exceed certain threshold and in case of the single linkage algorithm it is
identical to the minimal spanning tree~\cite{Gower1969}.

\begin{figure}[top]
\begin{center}
\includegraphics[width=7.0cm]{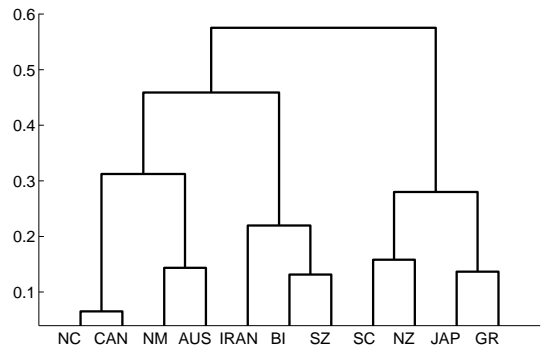}
\end{center}
\caption{
 Dendrogram representing the hierarchical clustering of the different earthquake networks for different regions, based on
the similarity of their spectral distributions. The distance metric $D$ is computed from the Jensen-Shannon divergence
between the corresponding spectral distributions of a pair of earthquake networks. Different spectral distributions are
obtained by convolving with a Gaussian kernel for $\sigma = 0.005$. The earthquake networks are clustered using complete
linkage algorithm. The height of the branch represents the distance between clusters, known as the linkage 
function $d$. 
The clusters are mainly formed  on the basis of their positions within different tectonic plates.}
\label{figure3}
\end{figure}

We use the complete linkage clustering algorithm to cluster the earthquake
networks hierarchically as shown in the dendrogram in Fig. \ref{figure3}.
We observe that different pairs of earthquake networks form clusters at the lower level. This pair-wise clustering of
earthquake networks reflects their relative positions in tectonic plates. The hierarchical clustering
indicates that the strongest structural similarity is observed between the earthquake networks of NC and CAN. A possible reason for this is that
NC and CAN belong to the North-American tectonic plate. Similarly, the clusters SZ-BI and JAP-GR belong to 
Eurasian tectonic plate. The earthquake network for SC is over the region close to the border of the Asia pacific 
and North American tectonic plate and NZ is over the Asia pacific plate. So, here we observe that SC-NZ together form a cluster.

 In the case of the underlying undirected structure of earthquake networks, we observe that their similarities between their 
 topologies are dependent on that between their underlying tectonic plates.
However, the directed version of each earthquake networks is strongly connected, implying that there is a directed chain of events from
one epicenter to an other. The strongly connected structure of earthquake networks motivates us to further analyse the directed version of the network. 
In our subsequent analysis, we focus on the directed chain of consecutive occurrence of earthquake events.

\begin{table*}[top]
\begin {center}
\begin{tabular}{c c c c c c c c c c c c} \hline
  Network  & SC        & NC       & JAP        & CAN       & IRAN      & GR        & NM        & BI     &  AUS   & SZ     & NZ     \\ \hline
  SC       & 0	       & 0.229    & 0.147      & 0.246     & 0.426     & 0.228     & 0.439     & 0.469  & 0.400  & 0.507  & 0.158 \\
  NC       & 0.229     & 0        & 0.316      & 0.065     & 0.382     & 0.314     & 0.312     & 0.391  & 0.290  & 0.459  & 0.338 \\
  JAP      & 0.147     & 0.316    & 0	       & 0.327     & 0.439     & 0.137     & 0.514     & 0.517  & 0.462  & 0.534  & 0.180 \\
  CAN      & 0.246     & 0.065    & 0.327      & 0	   & 0.357     & 0.320     & 0.289     & 0.368  & 0.261  & 0.438  & 0.360 \\
  IRAN     & 0.426     & 0.382	  & 0.439      & 0.357     & 0	       & 0.397     & 0.294     & 0.198  & 0.199  & 0.220  & 0.503 \\
  GR       & 0.228     & 0.314    & 0.137      & 0.320     & 0.397     & 0	   & 0.509     & 0.499  & 0.447  & 0.510  & 0.280 \\
  NM       & 0.439     & 0.312    & 0.514      & 0.289     & 0.294     & 0.509     & 0	       & 0.222  & 0.143  & 0.321  & 0.525 \\
  BI       & 0.469     & 0.391    & 0.517      & 0.368     & 0.198     & 0.499     & 0.222     & 0	& 0.157  & 0.132  & 0.548 \\
  AUS      & 0.400     & 0.290    & 0.462      & 0.261     & 0.199     & 0.447     & 0.144     & 0.157  & 0	 & 0.257  & 0.492 \\ 
  SZ       & 0.507     & 0.459    & 0.534      & 0.438     & 0.220     & 0.510     & 0.321     & 0.132  & 0.257  & 0	  & 0.575 \\ 
  NZ       & 0.158     & 0.338    & 0.180      & 0.360     & 0.503     & 0.280     & 0.525     & 0.548  & 0.492  & 0.575  & 0     \\  \hline
\end{tabular}
\end {center}
\caption{Structural distance table between earthquake networks using our method.
}\label{table2}
\end {table*}

\begin{table*}[top] 
\begin {center}
\begin{tabular}{|l|l|l|l|l|l|} \hline

  Region       & $\theta, \phi, m $   & $\theta, \phi, m $   & $\theta, \phi, m $   & $\theta, \phi, m $   &  $\theta, \phi, m $  \\ \hline

  SC           & 34.36, -116.45, 7815 & 35.05, -117.70, 6123 & 35.74, -117.63, 5467 & 36.03, -117.77, 5419 & 33.50, -116.45, 5291 \\

               & 33.67, -116.73, 5259 & 36.09, -117.84, 4865 & 36.03, -117.84, 4689 & 35.80, -117.63, 4440 & 33.50, -116.52, 4432 \\ \hline

  NC           & 37.56, -118.43, 3935 & 38.78, -122.75, 3598 & 38.78, -122.84, 3444 & 38.85, -122.84, 3001 & 37.63, -118.89, 2484 \\       

               & 37.56, -118.80, 2332 & 37.56, -118.89, 2290 & 37.28, -121.65, 2140 & 38.85, -122.75, 2085 & 37.49, -118.43, 1933 \\ \hline

  JAP          & 34.97, 139.16, 2026  & 34.18, 135.25, 1316  & 42.02, 139.16, 1254  & 36.14, 139.90, 1207  & 35.75, 137.45, 1158 \\

               & 36.73, 139.41, 1133  & 33.99, 135.25, 1080  & 35.55, 140.14, 1035  & 32.03, 130.35, 1023  & 36.14, 140.14, 997  \\ \hline

  CAN          & 47.39, -70.08, 3412  & 52.48, -131.70, 2834 & 52.94, -132.70, 2520 & 50.63, -130.71, 1481 & 48.78, -122.76, 1385 \\

               & 49.71, -127.73, 1366 & 50.17, -127.73, 1172 & 47.86, -70.08, 1121  & 48.78, -128.72, 1116 & 48.32, -122.76, 1087 \\  \hline

  IRAN         & 38.40, 46.73, 173    & 30.84, 56.85, 105    & 27.67, 56.56, 85     & 28.41, 51.64, 84     & 33.77, 48.75, 83     \\

               & 35.72, 49.04, 81     & 32.55, 48.75, 72     & 28.41, 57.14, 70     & 33.77, 49.04, 69     & 27.43, 55.98, 68      \\  \hline                  

  GR           & 38.39, 21.89, 2774   & 38.39, 22.06, 2086   & 37.23, 22.06, 1437   & 38.26, 22.06, 1222   & 38.39, 21.72, 890    \\

               & 38.26, 21.72, 838    & 37.62, 20.90, 655    & 38.26, 22.22, 607    & 38.39, 22.22, 595    & 38.01, 21.56, 582    \\  \hline              

  NM           & 36.56, -89.64, 1343  & 36.19, -89.42, 848   & 36.37, -89.42, 615   & 36.37, -89.64, 452   & 36.19, -89.64, 244   \\

               & 36.01, -89.87, 242   & 35.27, -92.35, 239   & 36.56, -89.42, 172   & 35.82, -90.09, 169   & 35.45, -92.35, 154   \\  \hline       

  BI           & 55.83, -3.09, 655    & 56.16, -3.69, 319    & 55.94, -3.09, 288    & 52.93, -4.49, 157    & 56.27, -3.69, 136    \\

               & 50.14, -5.09, 112    & 53.04, -2.29, 107    & 53.15, -1.10, 99     & 53.04, -2.09, 88     & 52.93, -4.29, 81     \\ \hline

  AUS          & -30.82, 117.24, 712  & -34.83, 149.16, 667  & -19.88, 134.02, 512  & -30.45, 117.24, 376  & -31.55, 116.83, 293  \\

               & -30.45, 117.65, 275  & -33.37, 138.52, 260  & -30.45, 116.83, 256  & -31.55, 117.24, 256  & -32.28, 117.24, 233  \\ \hline

  SZ           & 46.32, 7.36, 184     & 46.32, 7.41, 181     & 47.58, 7.61, 99      & 46.32, 7.32, 83      & 45.99, 7.90, 73      \\

               & 46.32, 7.46, 71      & 47.68, 7.46, 68      & 46.55, 10.31, 68     & 46.29, 7.22, 66      & 46.35, 7.41, 62      \\ \hline

  NZ           & -39.03, 175.15, 13417 & -45.14, 167.08, 13127 & -38.36, 176.04, 11752 & -40.39, 176.04, 11642 & -39.71, 176.94, 11385 \\

               & -37.68, 176.94, 11327 & -41.07, 175.15, 10664 & -39.03, 176.04, 9916  & -41.75, 174.25, 9861  & -43.79, 172.46, 8070  \\ \hline

\end{tabular}
\end {center}
\caption{Ten major earthquake prone locations are identified for different earthquake regions. 
Here, $\theta$ and $\phi$ represents the latitude and longitude in degree of the midpoint of a square cell with side $\ell=0.01$.
The total number of earthquake events within the cell is denoted by $m$.}\label{table3}
\end {table*}
Next, we study the highly earthquake prone locations of different regions as it has been reported earlier~\cite{Abe1,Abhijit15} that the degree
distributions of unweighted as well as the strength distribution of weighted earthquake
networks are highly heterogeneous with  power-law distributions. It indicates that few nodes of the network are having very high degree or
strength known as hub and large number of nodes are having small degree or strength. We have found the exact location of the node
(the mid point of a square cell where maximum number of earthquakes occurred) having maximum strength $s_{max}$ in different earthquake regions. 
The exact latitude-longitude coordinates of the nodes having maximum strength are (34.36, -116.45),(37.56, -118.43), (34.97, 139.16) (47.39, -70.08), 
(38.40, 46.73), (38.39, 21.89), (36.56, -89.64), (55.83, -3.09), (-30.82, 117.24), (46.32, 7.36), (-39.03, 175.15) with
$s_{max}$ = 7815, 3935, 2026, 3412, 173, 2774, 1343, 655, 712, 184, 13417  for SC, NC, JAP, CAN, IRAN, GR, NM, BI, AUS, SZ and NZ, respectively.
Top ten earthquake prone locations of each earthquake region are shown in the Table \ref{table3}.

\subsection{Part B: Statistical analysis of earthquake regions}
Now, we analyze the consecutive occurrence of earthquake events.
\subsubsection{Number of consecutive earthquakes}
  First, we quantify the pair of nodes having the link with maximum weight $w_{max}$ for different earthquake regions. Here, the maximum weight
  indicates highest number of consecutive earthquake events between the pair of nodes in that earthquake region. The exact latitude-longitude
  coordinates for the top ten pairs of nodes with maximum weight $w_{AB}$ are shown in the third column of the Table \ref{table4}-\ref{table5}.

  Now, using the same column ($w_{AB}$) of the above mention tables, we may roughly cluster the earthquake regions
  into four clusters consisting of (i) SC, NZ, NC
  (ii) IRAN, SZ (iii) CAN, NM, JAP (iv) GR, BI, AUS. In this clustering, most of the regions that are geographically proximal or close in the 
  dendrogram (Fig. 3) appear in the same cluster.
  
\begin{table}[top]
\begin {center}
\begin{tabular}{|c|c|c|c|c|c|c|c|c|} \hline
  $A$       &     $B$      & $w_{AB}$          & P($B^*\mid A^*$)   & $Z_{B^*|A^*}$\\ \hline
  \hline
  \multicolumn{5}{|c|}{SC}\\
\hline
35.74,     	    -117.63 	    & 	   35.80,    	    -117.63   	  &  	 1111  & 	 0.20 	 & 154.06\\
35.80,     	    -117.63 	    & 	   35.74,    	    -117.63   	  &  	 1066  & 	 0.24 	 & 147.69\\
36.03,     	    -117.91 	    & 	   36.03,    	    -117.84   	  &  	 552  & 	 0.15 	 & 87.58\\
36.03,     	    -117.84 	    & 	   36.03,    	    -117.91   	  &  	 545  & 	 0.12 	 & 86.32\\
34.02,     	    -116.31 	    & 	   33.96,    	    -116.31   	  &  	 422  & 	 0.11 	 & 92.21\\
33.96,     	    -116.31 	    & 	   34.02,    	    -116.31   	  &  	 401  & 	 0.16 	 & 87.51\\
33.16,     	    -115.61 	    & 	   33.21,    	    -115.61   	  &  	 387  & 	 0.11 	 & 96.99\\
33.04,     	    -114.99 	    & 	   35.05,    	    -117.70   	  &  	 380  & 	 0.21 	 & 76.06\\
33.21,     	    -115.61 	    & 	   33.16,    	    -115.61   	  &  	 380  & 	 0.18 	 & 95.31\\
33.04,     	    -114.99 	    & 	   35.05,    	    -117.70   	  &  	 380  & 	 0.21 	 & 76.06\\ 
\hline
  \multicolumn{5}{|c|}{NC}\\
  \hline
37.63,     	    -118.43 	    & 	   37.56,    	    -118.43   	  &  	 530  & 	 0.32 	 & 52.80\\
37.56,     	    -118.43 	    & 	   37.63,    	    -118.43   	  &  	 509  & 	 0.13 	 & 49.66\\
37.49,     	    -118.43 	    & 	   37.56,    	    -118.43   	  &  	 450  & 	 0.23 	 & 39.02\\
37.56,     	    -118.43 	    & 	   37.49,    	    -118.34   	  &  	 448  & 	 0.11 	 & 39.19\\
37.56,     	    -118.43 	    & 	   37.49,    	    -118.43   	  &  	 442  & 	 0.11 	 & 37.70\\
37.49,     	    -118.34 	    & 	   37.56,    	    -118.43   	  &  	 406  & 	 0.22 	 & 35.07\\
37.49,     	    -118.62 	    & 	   37.42,    	    -118.62   	  &  	 319  & 	 0.21 	 & 54.96\\
37.42,     	    -118.62 	    & 	   37.49,    	    -118.62   	  &  	 296  & 	 0.19 	 & 50.61\\
38.78,     	    -122.84 	    & 	   38.78,    	    -122.75   	  &  	 247  & 	 0.07 	 & 8.57\\
37.49,     	    -118.43 	    & 	   37.63,    	    -118.43   	  &  	 239  & 	 0.12 	 & 33.00\\
\hline
\multicolumn{5}{|c|}{JAP}\\
\hline
34.97,     	    139.16 	    & 	   34.97,    	    139.41   	  &  	 206  & 	 0.10 	 & 92.04\\
34.97,     	    139.41 	    & 	   34.97,    	    139.16   	  &  	 182  & 	 0.41 	 & 81.40\\
34.57,     	    135.00 	    & 	   34.77,    	    135.25   	  &  	 128  & 	 0.15 	 & 68.33\\
34.77,     	    135.25 	    & 	   34.57,    	    135.00   	  &  	 126  & 	 0.17 	 & 67.26\\
42.02,     	    139.16 	    & 	   41.82,    	    139.16   	  &  	 70  & 	 0.06 	 & 27.95\\
34.77,     	    139.41 	    & 	   34.97,    	    139.16   	  &  	 66  & 	 0.19 	 & 32.50\\
34.97,     	    139.16 	    & 	   34.77,    	    139.41   	  &  	 59  & 	 0.03 	 & 28.72\\
41.82,     	    139.16 	    & 	   42.02,    	    139.16   	  &  	 54  & 	 0.07 	 & 21.06\\
42.02,     	    139.41 	    & 	   42.02,    	    139.16   	  &  	 48  & 	 0.06 	 & 18.73\\
34.57,     	    139.41 	    & 	   34.77,    	    139.41   	  &  	 47  & 	 0.21 	 & 72.62\\
\hline
\multicolumn{5}{|c|}{CAN}\\
\hline
52.48,     	    -131.70 	    & 	   52.94,    	    -132.70   	  &  	 284  & 	 0.10 	 & 19.21\\
52.94,     	    -132.70 	    & 	   52.48,    	    -131.70   	  &  	 262  & 	 0.10 	 & 16.98\\
52.48,     	    -132.70 	    & 	   52.48,    	    -131.70   	  &  	 184  & 	 0.18 	 & 23.04\\
52.48,     	    -131.70 	    & 	   52.48,    	    -132.70   	  &  	 173  & 	 0.06 	 & 21.01\\
52.94,     	    -132.70 	    & 	   52.48,    	    -132.70   	  &  	 158  & 	 0.06 	 & 20.51\\
52.48,     	    -132.70 	    & 	   52.94,    	    -132.70   	  &  	 145  & 	 0.14 	 & 18.52\\
52.48,     	    -131.70 	    & 	   47.39,    	    -70.08   	  &  	 107  & 	 0.04 	 & -2.23\\
52.94,     	    -132.70 	    & 	   47.39,    	    -70.08   	  &  	 100  & 	 0.04 	 & -1.65\\
47.39,     	    -70.08 	    & 	   52.94,    	    -132.70   	  &  	 92  & 	 0.03 	 & -2.39\\
52.02,     	    -131.70 	    & 	   52.48,    	    -131.70   	  &  	 90  & 	 0.09 	 & 7.84\\
 \hline
  \multicolumn{5}{|c|}{IRAN}\\
\hline
38.40,     	    46.73 	    & 	   38.65,    	    46.73   	  &  	 18  & 	 0.10 	 & 31.02\\
38.65,     	    46.73 	    & 	   38.40,    	    46.73   	  &  	 17  & 	 0.47 	 & 29.37\\
28.41,     	    59.17 	    & 	   28.16,    	    59.17   	  &  	 15  & 	 0.42 	 & 45.17\\
28.16,     	    59.17 	    & 	   28.41,    	    59.17   	  &  	 15  & 	 0.26 	 & 45.14\\
28.41,     	    59.17 	    & 	   28.16,    	    59.17   	  &  	 15  & 	 0.42 	 & 45.17\\
28.16,     	    59.17 	    & 	   28.41,    	    59.17   	  &  	 15  & 	 0.26 	 & 45.14\\
36.45,     	    51.64 	    & 	   36.45,    	    51.35   	  &  	 14  & 	 0.29 	 & 43.94\\
36.45,     	    51.35 	    & 	   36.45,    	    51.64   	  &  	 12  & 	 0.30 	 & 37.63\\
30.84,     	    56.56 	    & 	   30.84,    	    56.85   	  &  	 12  & 	 0.33 	 & 26.64\\
36.45,     	    51.35 	    & 	   36.45,    	    51.64   	  &  	 12  & 	 0.30 	 & 37.63\\
\hline
\multicolumn{5}{|c|}{GR}\\
\hline
38.39,     	    22.06 	    & 	   38.39,    	    21.89   	  &  	 192  & 	 0.09 	 & 22.19\\
38.39,     	    21.89 	    & 	   38.39,    	    22.06   	  &  	 187  & 	 0.07 	 & 21.38\\
38.26,     	    22.06 	    & 	   38.39,    	    22.06   	  &  	 69  & 	 0.06 	 & 11.16\\
38.13,     	    26.52 	    & 	   38.13,    	    26.68   	  &  	 65  & 	 0.20 	 & 71.54\\
38.39,     	    21.89 	    & 	   38.39,    	    21.72   	  &  	 62  & 	 0.02 	 & 9.82\\
38.39,     	    22.06 	    & 	   38.26,    	    22.06   	  &  	 62  & 	 0.03 	 & 9.54\\
38.39,     	    21.89 	    & 	   38.39,    	    21.72   	  &  	 62  & 	 0.02 	 & 9.82\\
38.39,     	    22.06 	    & 	   38.26,    	    22.06   	  &  	 62  & 	 0.03 	 & 9.54\\
37.62,     	    20.90 	    & 	   37.75,    	    20.90   	  &  	 59  & 	 0.09 	 & 37.56\\
38.39,     	    21.72 	    & 	   38.39,    	    21.89   	  &  	 57  & 	 0.06 	 & 8.74\\ 
\hline
\end{tabular}
\end {center}
\caption{Statistical analysis of earthquake regions for SC, NC, JAP, CAN, IRAN and GR. 
$w_{AB}$ represents the total number of successive earthquake events between node $A$ and $B$. 
$P(B^*|A^*)$ is the conditional probability that there will be an earthquake at $B^*$ immediately after one occurs at $A^*$.
$Z_{B^*|A^*}$ is the Z-score for the successive events $A^*B^*$.
}\label{table4}
\end {table}

\begin{table}[top]
\begin {center}
\begin{tabular}{|c|c|c|c|c|c|c|c|c|} \hline
  $A$       &     $B$      & $w_{AB}$          & P($B^*\mid A^*$)   & $Z_{B^*|A^*}$\\ \hline
  \hline
\multicolumn{5}{|c|}{NM}\\
\hline
36.56,     	    -89.64 	    & 	   36.19,    	    -89.42   	  &  	 213  & 	 0.16 	 & 5.61\\
36.19,     	    -89.42 	    & 	   36.56,    	    -89.64   	  &  	 197  & 	 0.23 	 & 4.37\\
36.37,     	    -89.42 	    & 	   36.56,    	    -89.64   	  &  	 165  & 	 0.27 	 & 6.07\\
36.56,     	    -89.64 	    & 	   36.37,    	    -89.42   	  &  	 148  & 	 0.11 	 & 4.04\\
36.56,     	    -89.64 	    & 	   36.37,    	    -89.64   	  &  	 121  & 	 0.09 	 & 4.84\\
36.37,     	    -89.64 	    & 	   36.56,    	    -89.64   	  &  	 118  & 	 0.26 	 & 4.80\\
36.19,     	    -89.42 	    & 	   36.37,    	    -89.42   	  &  	 106  & 	 0.12 	 & 4.80\\
36.37,     	    -89.42 	    & 	   36.19,    	    -89.42   	  &  	 83  & 	 0.13 	 & 1.93\\
36.56,     	    -89.64 	    & 	   36.01,    	    -89.87   	  &  	 68  & 	 0.05 	 & 4.00\\
36.37,     	    -89.64 	    & 	   36.19,    	    -89.42   	  &  	 62  & 	 0.14 	 & 1.80\\ 
\hline
  \multicolumn{5}{|c|}{BI} \\
  \hline
55.94,     	    -3.09 	    & 	   55.83,    	    -3.09   	  &  	 115  & 	 0.40 	 & 19.32\\
55.83,     	    -3.09 	    & 	   55.94,    	    -3.09   	  &  	 105  & 	 0.16 	 & 16.77\\
55.83,     	    -3.09 	    & 	   56.16,    	    -3.69   	  &  	 30  & 	 0.05 	 & 0.65\\
53.04,     	    -2.29 	    & 	   53.04,    	    -2.09   	  &  	 30  & 	 0.28 	 & 26.37\\
55.83,     	    -3.09 	    & 	   56.16,    	    -3.69   	  &  	 30  & 	 0.05 	 & 0.65\\
53.04,     	    -2.29 	    & 	   53.04,    	    -2.09   	  &  	 30  & 	 0.28 	 & 26.37\\
55.83,     	    -3.09 	    & 	   52.93,    	    -4.49   	  &  	 27  & 	 0.04 	 & 3.85\\
53.04,     	    -2.09 	    & 	   53.04,    	    -2.29   	  &  	 26  & 	 0.30 	 & 22.72\\
56.16,     	    -3.69 	    & 	   55.83,    	    -3.09   	  &  	 23  & 	 0.07 	 & -0.76\\
53.48,     	    -2.09 	    & 	   53.48,    	    -2.29   	  &  	 23  & 	 0.77 	 & 64.37\\ 
  \hline
  \multicolumn{5}{|c|}{AUS}\\
  \hline
-30.45,     	    117.24 	    & 	   -30.45,    	    116.83   	  &  	 77  & 	 0.20 	 & 32.18\\
-30.45,     	    116.83 	    & 	   -30.45,    	    117.24   	  &  	 74  & 	 0.29 	 & 30.94\\
-30.09,     	    117.65 	    & 	   -30.45,    	    117.65   	  &  	 51  & 	 0.50 	 & 41.28\\
-30.45,     	    117.65 	    & 	   -30.09,    	    117.65   	  &  	 45  & 	 0.16 	 & 36.11\\
-31.91,     	    116.83 	    & 	   -31.55,    	    116.83   	  &  	 33  & 	 0.21 	 & 19.67\\
-31.55,     	    116.83 	    & 	   -31.91,    	    116.83   	  &  	 32  & 	 0.11 	 & 18.97\\
-34.83,     	    149.16 	    & 	   -30.82,    	    117.24   	  &  	 30  & 	 0.04 	 & 1.02\\
-30.45,     	    117.24 	    & 	   -30.82,    	    117.24   	  &  	 29  & 	 0.08 	 & 4.05\\
-30.82,     	    117.24 	    & 	   -30.45,    	    117.24   	  &  	 28  & 	 0.04 	 & 3.75\\
-19.88,     	    134.02 	    & 	   -34.83,    	    149.16   	  &  	 23  & 	 0.04 	 & 1.21\\
\hline
\multicolumn{5}{|c|}{SZ}\\
\hline
46.29,     	    7.17 	    & 	   46.29,    	    7.22   	  &  	 22  & 	 0.37 	 & 40.92\\
46.29,     	    7.22 	    & 	   46.29,    	    7.17   	  &  	 17  & 	 0.26 	 & 31.48\\
47.35,     	    11.03 	    & 	   47.35,    	    10.99   	  &  	 12  & 	 0.63 	 & 70.96\\
47.58,     	    7.61 	    & 	   47.64,    	    7.56   	  &  	 11  & 	 0.11 	 & 32.35\\
47.35,     	    10.99 	    & 	   47.35,    	    11.03   	  &  	 11  & 	 0.52 	 & 65.03\\
47.58,     	    7.61 	    & 	   47.64,    	    7.56   	  &  	 11  & 	 0.11 	 & 32.35\\
47.35,     	    10.99 	    & 	   47.35,    	    11.03   	  &  	 11  & 	 0.52 	 & 65.03\\
46.32,     	    7.41 	    & 	   46.32,    	    7.36   	  &  	 10  & 	 0.06 	 & 4.97\\
47.68,     	    7.51 	    & 	   47.68,    	    7.46   	  &  	 10  & 	 0.33 	 & 25.88\\
47.58,     	    7.61 	    & 	   47.61,    	    7.56   	  &  	 10  & 	 0.10 	 & 31.46\\ 
\hline
\multicolumn{5}{|c|}{NZ}\\
\hline
-45.14,     	    167.08 	    & 	   -39.03,    	    175.15   	  &  	 725  & 	 0.06 	 & 3.39\\
-37.68,     	    176.94 	    & 	   -45.14,    	    167.08   	  &  	 692  & 	 0.06 	 & 6.64\\
-39.03,     	    175.15 	    & 	   -45.14,    	    167.08   	  &  	 681  & 	 0.05 	 & 1.61\\
-39.71,     	    176.94 	    & 	   -40.39,    	    176.04   	  &  	 679  & 	 0.06 	 & 9.14\\
-39.03,     	    175.15 	    & 	   -38.36,    	    176.04   	  &  	 642  & 	 0.05 	 & 2.90\\
-40.39,     	    176.04 	    & 	   -39.03,    	    175.15   	  &  	 637  & 	 0.05 	 & 2.94\\
-45.14,     	    167.08 	    & 	   -37.68,    	    176.94   	  &  	 633  & 	 0.05 	 & 4.02\\
-40.39,     	    176.04 	    & 	   -39.71,    	    176.94   	  &  	 629  & 	 0.05 	 & 6.81\\
-41.07,     	    175.15 	    & 	   -39.03,    	    175.15   	  &  	 616  & 	 0.06 	 & 4.27\\
-38.36,     	    176.04 	    & 	   -39.03,    	    175.15   	  &  	 615  & 	 0.05 	 & 1.75\\
 \hline
\end{tabular}
\end {center}
\caption{Statistical analysis of earthquake regions for NM, BI, AUS, SZ and NZ. 
$w_{AB}$ represents the total number of successive earthquake events between node $A$ and $B$.  
$P(B^*|A^*)$ is the conditional probability that there will be an earthquake at $B^*$ immediately after one occurs at $A^*$.
$Z_{B^*|A^*}$ is the Z-score for the successive events $A^*B^*$.}\label{table5}
\end {table}
 
 \subsubsection{Conditional probability in earthquake networks} 
 Now we study the conditional probability of the occurrence of an earthquake event at node $B$, right after one takes place at node $A$.
 
 Let $A^*$ denote an earthquake at epicenter $A$ and $B^*$ is the same for $B$. The conditional probability of $B^*$ given $A^*$ is : 
 \begin{equation}
 P(B^*\mid A^*)=\frac{P(A^*\cap B^*)}{P(A^*)}.
 \end{equation}
 where $P(A^*\cap B^*)= w_{AB}/(n - 1)$  represents probability that the earthquake happened at $A$ and $B$ consequently and $P(A^*)=s_A/n$
 is the probability of earthquake events at $A^*$. 
Here we compute the conditional probability between the top ten pairs of nodes having maximum link weight for each earthquake region.

The exact latitude-longitude coordinates of the top 10 two consecutive nodes are given in Table \ref{table4}-\ref{table5} (see the fourth column).

Here we see that $P(B^* | A^*)$, in all the earthquake regions, for some $A$ and $B$ are not negligible. The higher values of $P(B^* | A^*)$
may depend on the high frequency of $B^*$. So to predict the occurrence of $B^*$ right after $A^*$, with high statistically significance, we compute the Z-score
for $P(B^* | A^*)$.
 
 \subsubsection{Statistical significance factor for consecutive events}   

The statistical significance can be measured by computing the Z-score for the successive events at $A$ and $B$:
\begin{equation}
  Z_{B^*|A^*} = \frac{P(B^*|A^*)-\pi(A^*, ~B^*)}{\sqrt\frac{{\pi(A^*, ~B^*)(1-\pi(A^*,~ B^*))}}{n(A^*)}}
\end{equation}
where $\pi$ is the stationary probability matrix (in our case, stationary matrix $\pi$ exists, since the underlying graph is connected, undirected and non-bipartite),
calculated from matrix $M$. The rows and columns of $M$ are different events and $(A^*, B^*)$-th element of $M$  is defined as:
$$M_{A^*B^*}=P(B^*|A^*),$$

 and $\sqrt\frac{{\pi(A^*,~B^*)(1-\pi(A^*,~B^*))}}{n(A^*)}$ is the corresponding standard deviation. Here, $n(A^*)$ denotes the cardinality of $A^*$.

We find Z-score values among top 10 two consecutive nodes for each of the region. This is shown in the fifth column of the Table \ref{table4}-\ref{table5}.
%

\section{Conclusion}
We have analyzed eleven earthquake catalogs from the different parts of the world using a complex network framework and statistical techniques. 
The graph spectra of different earthquake regions appear to be very different from the spectrum of Erd\H os-R\'enny random network. The spectral probability
functions of various regions have distinct peakedness. The dissimilarity between the spectral probability functions are shown as a distance
matrix, which is calculated using a probabilistic metric, viz., Jensen-Shannon divergence. We have hierarchically clustered the earthquake regions from
the distance matrix and linked it with their proximity in tectonic plates. The locations of highly earthquake prone regions have also been identified. 
To understand the chain of earthquake events, we have delved deeper by considering directed version of these networks, and it is revealed that the entire network is strongly
connected for all regions. Furthermore, we have calculated the conditional probability of a forthcoming earthquake event on a node and the statistical significance
of those probabilities is also estimated by comparing with their stationary probability. 

\noindent
\begin{center}
{\bf ACKNOWLEDGMENTS}
\end{center}
We thank Shakti N. Menon for helpful suggestions.
We gratefully acknowledge the assistance of Buddhananda Banerjee in the statistical part of this work. 
KD gratefully acknowledge the financial support from CSIR, India.


\begin{thebibliography}{90}
\bibitem {GR} B. B. Gutenberg and C. F. Richter, Bull. Seismol. Soc. Am. {\bf 34}, 185 (1944).
\bibitem {Omori} F. Omori, J. Coll. Sci. Imp. Univ. Tokyo {\bf 7}, 111 (1895).
\bibitem {BK} R. Burridge and L. Knopoff, Bull. Seismol. Soc. Am. {\bf 57}, 341 (1967).
\bibitem{Cox09} A. Cox and R. B. Hart, {\em Plate tectonics: how it works} (John Wiley \& Sons, 2009).
\bibitem{Bak} P. Bak, K. Christensen, L. Danon and T. Scanlon, Phys. Rev. Lett. {\bf 88}, 178501 (2002).
\bibitem{Baiesi2004} M. Baiesi and M. Paczusk, Phys. Rev. E {\bf 69}, 066106 (2004).
\bibitem{Davidsen2005} J. Davidsen and M. Paczuski, Phys. Rev. Lett. {\bf 94}, 048501 (2005).
\bibitem {Abe1} S. Abe and N. Suzuki, Euro. Phys. Lett. {\bf 65}, 581 (2004).
\bibitem {Abe2} S. Abe and N. Suzuki, Physica A {\bf 337}, 357 (2004).
\bibitem {Abe3} S. Abe and N. Suzuki, Phys. Rev. E {\bf 74}, 026113 (2006).
\bibitem {Abe4} S. Abe and N. Suzuki, Euro. Phys. J. B {\bf 59}, 93 (2007).
\bibitem {Abe5} S. Abe and N. Suzuki, Physica A {\bf 388}, 1917 (2009).
\bibitem {Abe6} S. Abe and N. Suzuki, Physica A {\bf 388}, 2511 (2009).
\bibitem {Abe7} S. Abe and N. Suzuki, Euro. Phys. Lett. {\bf 87}, 48008 (2009).
\bibitem {Abe8} S. Abe, D. Pasten and N. Suzuki, Physica A {\bf 390}, 1343 (2011).
\bibitem{Abhijit15} A. Chakraborty, G. Mukherjee and S.~S. Manna, Physica A {\bf 433}, 336 (2015).
\bibitem{Johnson1967} S.~C.~Johnson, Psychometrika {\bf 2}, 241 (1967).
\bibitem{Abe05} S. Abe and N. Suzuki, Euro. Phys. J. B {\bf 44} 1 (2005).
\bibitem{Abe9} S. Abe and N. Suzuki, Euro. Phys. Lett. {\bf 97}, 49002 (2012).
\bibitem{Endres2003}D. M. Endres and J. E. Schindelin, IEEE Trans. Inf. Theory {\bf 49}, 1858 (2003).
\bibitem{Lin1991} J. Lin, IEEE Trans. Inf. Theory {\bf 37}, 145 (1991).
\bibitem {Anirban} A. Banerjee, BioSystems {\bf 107}, 186 (2012).
\bibitem {Deyasi15} K. Deyasi, A. Banerjee and B. Deb, J. Biosci. {\bf 40}, 799 (2015).
\bibitem{BanerjeeJost2008a} A. Banerjee and J. Jost, Linear Algebra Appl. {\bf 428}, 3015 (2008).
\bibitem{BanerjeeJost2009} A. Banerjee and J. Jost, Discrete Appl. Math. {\bf 157}, 2425 (2009).
\bibitem{Lange2014} S.~C. de Lange , M.~A. de Reus and M.~P. van den Heuvel, Front Comput Neurosci {\bf 7}, 189 (2014). 
\bibitem{Chung} F.~R.~K. Chung, {\em Lecture notes on spectral graph theory} (AMS Publications Providence, 1997).
\bibitem{Gower1969} J.~C. Gower and G.~J.~ S. Ross, J. Roy. Stat. Soc. C {\bf18}, 54 (1969).
\end{thebibliography}
\end {document}